\documentstyle[12pt,epsfig]{article}
\newcommand{\Frac}[2]%
{{\textstyle \frac{\mbox{\footnotesize $#1$}\rule[-0.9mm]{0mm}{1mm}}%
{\mbox{\footnotesize $#2$}\rule{0mm}{3.1mm}}}}
\renewcommand{\thefootnote}{\fnsymbol{footnote}}

\begin{document}
\begin{titlepage}
\vspace*{-12 mm}
\noindent
\begin{flushright}
\begin{tabular}{l@{}}
\end{tabular}
\end{flushright}
\vskip 12 mm
\begin{center}
{\large \bf 
Anomalous commutators and electroweak baryogenesis
%\\[3mm]  
}
\\[14 mm]
{\bf Steven D. Bass}
\\[10mm]   
{\em High Energy Physics Group, \\
Institute for Experimental Physics and
Institute for Theoretical Physics, \\
Universit\"at Innsbruck, \\
Technikerstrasse 25, A 6020 Innsbruck, Austria
\\[5mm]
}
\end{center}
\vskip 10 mm
\begin{abstract}
\noindent
Electroweak vacuum transition processes (sphalerons) in the early 
Universe provide a possible explanation of the baryon asymmetry. 
Combining this physics with the anomalous commutators of Adler and 
Boulware and renormalization group invariance, 
we argue that electroweak baryon number violation also induces a
``topological condensate'' in the vacuum.
QCD sphaleron processes act to distribute the baryon number violation
between both left and right handed quarks and induce a spin independent
component in this ``condensate''.
\end{abstract}
\end{titlepage}
\renewcommand{\labelenumi}{(\alph{enumi})}
\renewcommand{\labelenumii}{(\roman{enumii})}
\renewcommand{\thefootnote}{\arabic{footnote}}
\newpage
\baselineskip=6truemm

\section{Introduction}

Electroweak baryogenesis offers the intriguing possibility that topological 
features of the Standard Model might be responsible for baryon and lepton 
number violation in the early Universe \cite{shaprev,shapplb,arnold}. 
In this paper we examine this physics in the context of the anomalous
commutator theory developed by Adler and Boulware \cite{boulware,brandeis}
together with renormalization group arguments. 
We argue that electroweak baryogenesis in the early Universe is accompanied 
by the formation of a ``topological condensate'' in the Standard Model vacuum.
When the effect of
QCD vacuum transition processes, which break axial U(1) symmetry, are 
also included the net ``topological condensate'' develops a spin independent
component.

We first outline the key features of electroweak baryogenesis and 
then explain the consequences of anomalous commutator theory for this physics.

In the Standard Model the parity violating SU(2) electroweak interaction 
induces an axial anomaly contribution \cite{adler} in the (vector) baryon 
number current \cite{thooft}.
Through electroweak instantons this leads to the possibilty of baryon 
(and lepton) number violation through quantum tunneling processes in
the $\theta$-vacuum for the Standard Model fields.
This baryon number violation never appears in perturbative calculations
but is generated through nonperturbative transitions between different
vacuum states.
Each transition violates baryon number (and lepton number)
by $\Delta B = \Delta L = \pm 3 n_f$
where $n_f$ is the number of families (or fermion generations).
For example, for $n_f=3$, we find electroweak processes such as
\begin{equation}
{\rm q \ + \ q \rightarrow 7 {\bar q} \ + \ 3 {\bar l}}
\label{eq1}
\end{equation}
where all the fermions in this equation are understood to be left handed
and there are 3 quarks and one lepton from each generation.
At zero temperature these transitions are exponentially suppressed
by the factor
\begin{equation}
e^{ - 4 \pi \sin^2 \theta_W / \alpha }
\sim 10^{-170}
\label{eq2}
\end{equation}
and are therefore entirely negligible.
Kuzmin, Rubakov and Shaposhnikov \cite{shapplb}
argued that this baryon number violating
process becomes unsuppressed at high temperature
$T \gg M_W$,
and is thus a candidate for baryon number violation in
the early Universe.
The reason that the suppression goes away is that the transition can
then arise due to thermal fluctuations rather than quantum tunneling
once the temperature becomes high compared to the potential barrier
$V_0$ between the different vacuum states.
\footnote{
The anomalous electroweak baryon number violating process might also be 
observable in (very) high energy proton-proton collisions 
when the centre of mass energy in the parton parton collision 
exceeds the potential barrier between the different vacuum states
\cite{farrar}.
}
These electroweak vacuum transitions involve (just) left-handed fermions
through the parity violating couplings of the electroweak SU(2) gauge fields.
Additional QCD sphaleron transition processes 
(mediated through the strong QCD axial anomaly) 
plus couplings to the scalar Higgs field(s) offer possible 
mechanisms for transfer of the baryon number violation to right handed
quarks \cite{shaprev}.

\section{The axial anomaly and anomalous commutators}

The vector baryon current can be written as the sum of left and right
handed currents:
\begin{equation}
J_{\mu}
= {\bar \Psi} \gamma_{\mu} \Psi
= {\bar \Psi} \gamma_{\mu} {1 \over 2} (1 - \gamma_5) \Psi
+ {\bar \Psi} \gamma_{\mu} {1 \over 2} (1 + \gamma_5) \Psi
.
\label{eq3}
\end{equation}
Classically, this fermion current is conserved.
However, in the Standard Model the left handed fermions 
couple to the SU(2) electroweak gauge fields $W$ and $Z^0$.
As 't Hooft first pointed out \cite{thooft}, this means 
that this baryon current is sensitive to the axial anomaly.
One finds the anomalous divergence equation
\begin{equation}
\partial^\mu J_{\mu}
= n_f
\biggl( - \partial^\mu K_\mu + \partial^\mu k_\mu
\biggr)
\label{eq4}
\end{equation}
where 
$K_{\mu}$ and $k_{\mu}$ are the 
SU(2) electroweak and U(1) hypercharge Chern-Simons currents
\begin{equation}
K_{\mu} = {g^2 \over 16 \pi^2}
\epsilon_{\mu \nu \rho \sigma}
\biggl[ A^{\nu}_a \biggl( \partial^{\rho} A^{\sigma}_a 
- {1 \over 3} g 
f_{abc} A^{\rho}_b A^{\sigma}_c \biggr) \biggr]
\label{eq5}
\end{equation}
and
\begin{equation}
k_{\mu} = {g'^2 \over 16 \pi^2}
\epsilon_{\mu \nu \rho \sigma} B^{\nu} \partial^{\rho} B^{\sigma}
.
\label{eq6}
\end{equation}
Here
$A_{\mu}$ and $B_{\mu}$ denote the SU(2) and U(1) gauge fields, and
$
\partial^{\mu} K_{\mu} 
= {g^2 \over 32 \pi^2} W_{\mu \nu} {\tilde W}^{\mu \nu}
$
and
$
\partial^{\mu} k_{\mu} 
= {g'^2 \over 32 \pi^2} F_{\mu \nu} {\tilde F}^{\mu \nu}
$
are the SU(2) and U(1) topological charge densities.
Eq.(\ref{eq4}) allows us to define a conserved current
\begin{equation}
J_{\mu }^{\rm con} = J_{\mu } - n_f (- K_{\mu} + k_{\mu})
.
\label{eq7}
\end{equation}
The current $J_{\mu }^{\rm con}$ satisfies the divergence equation
\begin{equation}
\partial^\mu J^{\rm con}_{\mu} = 0
\label{eq8}
\end{equation}
but is SU(2) and U(1) gauge dependent because of the gauge dependence of 
$K_{\mu}$ and $k_{\mu}$.
When we make a gauge transformation $U$ the SU(2) electroweak gauge 
field transforms as
\begin{equation}
A_{\mu} \rightarrow U A_{\mu} U^{-1} + {i \over g} (\partial_{\mu} U) U^{-1}
\label{eq10}
\end{equation}
and the operator $K_{\mu}$ transforms as
\begin{equation}
K_{\mu} \rightarrow K_{\mu} 
+ i {g \over 8 \pi^2} \epsilon_{\mu \nu \alpha \beta}
\partial^{\nu} 
\biggl( U^{\dagger} \partial^{\alpha} U A^{\beta} \biggr)
+ {1 \over 24 \pi^2} \epsilon_{\mu \nu \alpha \beta}
\biggl[ 
(U^{\dagger} \partial^{\nu} U) 
(U^{\dagger} \partial^{\alpha} U)
(U^{\dagger} \partial^{\beta} U) 
\biggr]
\label{eq11}
\end{equation}
where the third term on the RHS is associated with the gauge field topology
\cite{crewther}.
Because of the absence of topological structure in the U(1) sector it is 
sufficient to drop the U(1) ``$k_{\mu}$ contribution'' in discussions of 
anomalous baryon number violation, which we do in all discussion below.
Conserved and partially conserved currents are not renormalized. 
It follows that $J^{\rm con}_{\mu}$ is renormalization scale invariant.
The gauge invariantly renormalized current $J_{\mu}$ 
is scale dependent with the two-loop anomalous dimension 
induced by the axial anomaly
--
the scale dependence of $J_{\mu}$ is carried entirely by $K_{\mu}$
\cite{adler,crewther}.

Equation (\ref{eq4}) presents us with two candidate currents we might 
try to use 
to define the baryon number:
$J_{\mu}$
and 
$J_{\mu}^{\rm con}$.
We next 
explain how both currents yield gauge invariant possible definitions. 
The selection which current to use has interesting physical consequences
which we discuss in Section 3.

We choose the $A_0$ (and $B_0=0$) gauge and define two operator charges:
\begin{equation}
Y(t) = \int d^3z \ J_{0} (z), \ \ \ \ \ \ \ \ \ \
B  = \int d^3z \ J_{0}^{\rm con}(z)
.
\label{eq12}
\end{equation}
Because conserved currents are not renormalized it follows that $B$ 
is a time independent operator.
The charge $Y(t)$ is manifestly gauge invariant whereas $B$ is invariant 
only under ``small'' gauge transformations; the charge $B$ transforms as
\begin{equation}
B \rightarrow B + n_f \ m
\label{eq15}
\end{equation}
where $m$ is the winding number associated with the gauge transformation $U$.
Although $B$ is gauge dependent we can define a gauge invariant 
baryon number ${\cal B}$ for a given operator ${\cal O}$ 
through the gauge-invariant eigenvalues of the equal-time commutator
\begin{equation}
[ \ B \ , \ {\cal O} \ ]_{-} = {\cal B} \ {\cal O} 
.
\label{eq17}
\end{equation}
(The gauge invariance of ${\cal B}$ follows since this commutator
 appears in gauge invariant Ward Identities \cite{crewther}, despite
 the gauge dependence of $B$.)
The time derivative of spatial components of the W-boson field 
have zero baryon number ${\cal B}$ but non-zero $Y$ charge:
\begin{equation}
[ \ B \ , \ \partial_0 A_i \ ]_- \ = \ 0
\label{eq19}
\end{equation}
and
\begin{equation}
\lim_{t' \rightarrow t}
\biggl[ \ Y(t') \ , \ \partial_0 A_i ({\vec x}, t) \ \biggr]_- 
\ = \ 
{i n_f g^2 \over 4 \pi^2} {\tilde W}_{0i} \ + \ O(g^4 \ln | t'-t| )
\label{eq20}
\end{equation}
with 
${\tilde W}_{\mu \nu}$ the SU(2) dual field tensor
(see Refs.\cite{boulware,brandeis,crewther} for a discussion of
 the analogous situation in QED and QCD)
.
Eq.(\ref{eq19}) follows from the non-renormalization of the conserved
current $J_{\mu}^{\rm con}$. 
Eq.(\ref{eq20}) follows from the implicit 
$A_{\mu}$ dependence of the (anomalous) gauge invariant current $J_{\mu}$.
The higher-order terms $g^4 \ln | t'-t|$ are caused by wavefunction
renormalization of $J_{\mu}$ \cite{crewther}.

Motivated by this discussion of anomalous commutators, 
plus the 
renormalization scale invariance of baryon number, 
we next choose to identify baryon (and lepton) number 
with the gauge invariant commutators of the charge $B$
associated with conserved current $J_{\mu}^{\rm con}$, Eq.(\ref{eq17}).
We investigate the physical consequences of this choice
\footnote{
Traditionally, the electroweak baryogenesis literature 
has implicitly assumed that baryon and lepton number is 
associated with the eigenvalues of 
the charge $Y$ of the gauge-invariantly renormalized current
$J_{\mu}$. 
We believe that the arguments presented above imply that the conserved
current definition presents at the minimum a legitimate alternative
whose physical consequences should be explored.
}
and compare our results with the physics obtained if one 
instead uses the gauge invariantly renormalized current 
$J_{\mu}$ and the charge $Y(t)$ to define the ``baryon number''.

Before proceeding further, it will also be helpful to introduce the 
axial-vector current
$
J_{\mu5} = \bar{\Psi }\gamma_\mu \gamma_5 \Psi
$
which is taken to be gauge-invariantly renormalized 
with the axial anomaly in the RHS of its divergence equation. 
Similar to our discussion above, and as standard in the axial
anomaly literature, we also introduce the gauge dependent but
partially conserved axial-vector current operator 
$J_{\mu 5}^{\rm con} = J_{\mu 5} - K_{\mu}|_{\rm QCD}$, 
with $K_{\mu}|_{\rm QCD}$ the QCD Chern-Simons current.
Under QCD gauge transformations
characterized by (QCD) winding number $n$
the charge
$Q_5  = \int d^3z \ J_{05}^{\rm con}(z)$
transforms as
$Q_5 \rightarrow Q_5 - 2 n$.
The partially conserved current is renormalization scale independent and
the commutators
$
[Q_5 , {\cal O}]_{-} = - {\cal Q}_5 \ {\cal O}
$
can be used to define a (QCD) gauge invariant chirality ${\cal Q}_5$.
The time derivative of spatial components of the gluon field have
vanishing $Q_5$ chirality and non-vanishing $X(t) = \int d^3z \ J_{05}(z)$ 
charge \cite{crewther}.

\section{Gauge topology and vacuum transition processes}

When topological effects are taken into account, 
the Standard Model vacuum $|\theta_1, \theta_2 \rangle$
is a coherent superposition 
\begin{equation}
|\theta_1, \theta_2 \rangle 
= \sum_m \sum_n {\rm e}^{i (m \theta_1 + n \theta_2)} 
|m \rangle_{\rm EW} \ |n \rangle_{\rm QCD}
\label{eq21}
\end{equation}
of the
eigenstates 
$|m \rangle_{\rm EW}$
of 
$\int d \sigma_{\mu} K^{\mu} \neq 0$
and
$|n \rangle_{\rm QCD}$ of 
$\int d \sigma_{\mu} K^{\mu}_{\rm QCD} \neq 0$ 
(with $K^{\mu}_{\rm QCD}$ the QCD Chern Simons current)
\cite{crewther}.
Here
$\sigma_{\mu}$ is a large surface which is defined 
such that its boundary is spacelike with respect to the positions 
$z_k$ of any operators or fields in the physical problem under discussion.
For integer values of the topological winding number $m$, 
the states 
$|m \rangle_{\rm EW}$ contain $4 m n_f$ fermions 
(3 quarks and one lepton from each fermion generation) 
carrying baryon and lepton number $B=L= n_f \xi_{\rm EW}$
(and zero net electric charge).
The factor $\xi_{\rm EW}$ is equal to +1 if the baryon number 
is associated with $J_{\mu }^{\rm con}$ and equal to -1 if the
baryon number is associated with $J_{\mu}$ --- see below.
Relative to the $|m=0 \rangle_{\rm EW}$ state, the $|m=+1 \rangle_{\rm EW}$ 
state carries electroweak topological winding number +1 and 
$3 n_f$ quarks and $n_f$ leptons 
with baryon and lepton number $B=L= n_f \xi_{\rm EW}$.
In the QCD part of the vacuum, for integer values of the QCD
topological winding number $n$,
the states $|n \rangle_{\rm QCD}$ 
contain $nf$ quark-antiquark pairs 
with non-zero $Q_5$ chirality
$\sum_l \chi_l = - 2 f n \xi_{\rm QCD}$ 
where $f$ is the number of light-quark flavours.
Relative to the $|n=0 \rangle_{\rm QCD}$ state, the $|n=+1 \rangle_{\rm QCD}$ 
state carries topological winding number +1 and $f$
quark-antiquark pairs with $Q_5$ chirality equal to $-2f \xi_{\rm QCD}$.
The factor $\xi_{\rm QCD}$ is equal to +1 if the $U_A(1)$ symmetry of 
QCD is associated with $J_{\mu 5}^{\rm con}$ and 
equal to -1 if the $U_A(1)$ symmetry is associated with $J_{\mu 5}$.

Following from Eqs.(\ref{eq4}) and (\ref{eq7}), in electroweak sphaleron 
(or instanton) induced vacuum transition processes
\begin{equation}
\Delta Y = \Delta B - n_f m
\label{eq23}
\end{equation}
where $m = \pm 1$ is the change in the electroweak topological winding number.
The change in winding number is an integer 
for these processes and renormalization scale independent.
The anomalous commutators (\ref{eq19},\ref{eq20}) and
renormalization group invariance 
suggest that we associate the change in the baryon number with 
the baryonic charge $B$ in this equation: $\Delta B = n_f m$ with $\Delta Y=0$.

We now consider the physical effect of the choice of baryon number current.
For the sake of definiteness we consider a vacuum transition
characterized by a change in the electroweak topological winding number $m=+1$.
\begin{enumerate}
\item
First consider the scenario where the baryon number is associated 
with the conserved vector current $J_{\mu}^{\rm con}$ through the
gauge invariant commutators of the charge $B$, Eq.(\ref{eq17}). 
Here
$\Delta B = n_f$ and $\Delta Y=0$ in the sphaleron transition process.
Energy and momentum are conserved 
between the particles which are produced and absorbed 
in the non-perturbative transition, eg. Eq.(\ref{eq1}), 
which produces the baryon and lepton number violation.
The topological term coupled to $K_{\mu}$ which measures the change in 
the winding number 
(or change in the gauge-field boundary conditions at infinity)
carries zero energy and zero momentum.
Thus, the change in the baryon number $\Delta B$ is compensated by a 
shift of quantum-numbers with equal magnitude but opposite sign into 
the ``vacuum''
(defined here as everything carrying zero four-momentum)
so that $Y$ is conserved.
In this scenario the ``vacuum'' acquires a
``topological charge''
which compensates the baryon and lepton number non-conservation
(plus chirality 
 since electroweak sphalerons/baryogenesis act just on left-handed fermions).
\item
In the alternative scenario where baryon number is identified with 
the current $J_{\mu}$
the non-conserved 
``baryon number'' is identified with $Y$ in Eq.(\ref{eq23}), 
viz. $\Delta Y = - n_f$ and $\Delta B=0$.
It is illuminating to consider the anatomy of this process, 
looking at 
the details needed to restore $B$ conservation.
For QCD instantons this was discussed by 't Hooft -- see Section 6 of
his paper \cite{thrept}.
An effective ``schizon'' object needs to be introduced to absorb in 
the ``vacuum'' $B$ quantum numbers equal in magnitude and 
opposite in sign to those induced by the change in the topological 
winding number.
The ``schizon'' carries zero energy and zero momentum and acts to cancel 
the zero-mode contribution obtained in 
the previous
``$J_{\mu}^{\rm con}$ baryon number'' scenario.
The ``schizon'' is constructed to produce a 
theory with no net transfer of quanta to the ``vacuum'' 
under instanton or sphaleron transition processes.
This contrasts with the first scenario where the vacuum does 
acquire net quantum numbers since the total $Y$ charge is conserved.
\end{enumerate}

What is the practical effect of 
using the charge $B$ to define baryon number ?

In this scenario the total charge $Y$ and the information measured by it 
are conserved in electroweak vacuum transitions:
the ``vacuum'' will acquire topological quantum numbers equal to minus the 
change in baryon (and lepton) number and minus the change in (left-handed) 
chirality induced by electroweak sphalerons.
That is, this electroweak baryogenesis process is accompanied by the 
formation of a ``topological condensate'' in the ``vacuum'' carrying 
these quantum numbers.
QCD sphalerons will also be at work in high temperature processes and
(together with Higgs couplings) 
act to distribute the baryon number violation between both left and 
right handed quarks \cite{shaprev}.
The same arguments that we used for baryon number violation 
generalize readily to axial U(1) violation in QCD instanton/sphaleron 
processes
\cite{bass98,crewther,thrept}.
It follows that, 
with $B \mapsto Q_5$ and $Y \mapsto X$, 
QCD sphalerons act to cancel the chirality dependence of the (quark 
part of the) electroweak vacuum generated by the electroweak sphalerons.
The quanta transfered to the ``vacuum'' ensure that both the total 
$X(t)$ and $Y(t)$ charges are conserved in instanton and sphaleron 
transition processes.
QCD sphaleron processes which shift the baryon number violation 
from left to right handed quarks also act to cancel the chiral 
polarization of the ``vacuum'' induced by electroweak sphalerons.
Hence, the result is to create a spin/chiral independent 
component in the net ``topological condensate'' which forms 
in electroweak baryogenesis 
(corresponding to finite baryon number violation with the chiral 
 dependence, at least in part, cancelled.)

The QCD analogue of this physics has been studied in the context of 
the proton spin and axial U(1) problems. 
For the proton spin problem, gluon topological effects 
have the potential to induce a ``subtraction at infinity'' 
correction to the Ellis-Jaffe sum-rule for polarized deep inelastic 
scattering \cite{zakopane}. 
This correction, if finite, corresponds to a Bjorken $x=0$ 
(or ``polarized condensate'' \cite{bass98})
contribution to the nucleon's flavour-singlet axial charge $g_A^{(0)}$
generated through dynamical axial U(1) symmetry breaking.
In the language of Regge phenomenology it is a $J=1$ fixed pole with 
nonpolynomial residue contribution to the spin dependent part of the 
real part of the forward Compton scattering amplitude.
A direct measurement of $g_A^{(0)}$,
independent of this possible correction,
could be obtained from a precision measurement of elastic $\nu p$ scattering.
Thus, the physics is amenable to experiment.

\section{Conclusions}

Arguments associated with anomalous commutator theory and renormalization 
group invariance suggest that sphaleron induced electroweak 
baryogenesis in the early Universe is accompanied by the formation of a 
``topological condensate''.
It seems reasonable to postulate that this ``topological condensate''
survives the cooling to present times along with the baryon number 
violation induced by baryogenesis.
QCD sphalerons tend to cancel the spin dependence 
of the quark part of 
this ``condensate'' leaving the baryon number violating component untouched.
The phenomenology 
of this ``condensate'' 
and possible implications for cosmology deserve further study.

\vspace{0.5cm}

{\bf Acknowledgements} \\

SDB thanks the Austrian Science Fund (FWF) for financial support (grant M770).
This work was begun at the 2004 Cracow Epiphany Conference on Astroparticle 
Physics. I thank L. McLerran for helpful communications.


\newpage


\begin{thebibliography}{99}
%
\bibitem{shaprev}
V.A. Rubakov and M.E. Shaposhnikov, Usp. Fiz. Nauk. {\bf 166}, 493 (1996);
{\tt hep-ph/9603208}.
%
\bibitem{shapplb}
V.A. Kuzmin, V.A. Rubakov and M.E. Shaposhnikov, Phys. Lett. {\bf B155}, 36
(1985).
%
\bibitem{arnold}
P. Arnold and L. McLerran, Phys. Rev. {\bf D36}, 581 (1987).
%
\bibitem{boulware}
S. L. Adler and D. G. Boulware, Phys. Rev. {\bf 184}, 1740 (1969).
%
\bibitem{brandeis}
S.L. Adler, in Brandeis 
{\it Lectures on Elementary Particles and Quantum Field Theory}, 
eds. S. Deser, M. Grisaru and H. Pendleton (MIT Press, 1970).
%
\bibitem{adler} 
S.L. Adler, Phys. Rev. {\bf 177}, 2426 (1969);
J.S. Bell and R. Jackiw, Nuovo Cimento {\bf 60A}, 47 (1969).
%
\bibitem{thooft}
G. 't Hooft, Phys. Rev. Lett. {\bf 37}, 8 (1976); 
             Phys. Rev. {\bf D14}, 3432 (1976).
%
\bibitem{farrar}
G.R. Farrar and R. Meng, Phys. Rev. Lett. {\bf 65}, 3377 (1990);
A. Ringwald, Nucl. Phys. {\bf B330}, 1 (1990).
%
\bibitem{crewther}
R.J. Crewther, 
Acta Physica Austriaca Suppl. {\bf 19}, 47 (1978).
%
\bibitem{thrept}
G. 't Hooft, Phys. Rept. {\bf 142}, 357 (1986).
%
\bibitem{bass98} 
S.D. Bass, Mod. Phys. Lett. {\bf A13}, 791 (1998).
%
\bibitem{zakopane}
S. D. Bass, Zakopane lectures, Acta Phys. Pol. B {\bf 34}, 5893 (2003).
%
\end{thebibliography}
\end{document}